\documentclass[aps,floats,prd,nofootinbib,twocolumn]{revtex4}

\usepackage{amssymb}
\usepackage{amsmath}

\usepackage{graphicx}
\usepackage{graphics}
\usepackage{dcolumn}
\usepackage{color}
\usepackage{fancyhdr} 
\usepackage{graphicx}

\usepackage{subfig}

\usepackage[utf8]{inputenc}

\usepackage[justification=centerlast, format=plain, labelfont=bf]{caption}

\def\VEV#1{\left\langle #1 \right\rangle}

    \newcommand{\be}{\begin{equation}}
  \newcommand{\ee}{\end{equation}}
    \newcommand{\ba}{\begin{align}}
  \newcommand{\ea}{\end{align}}

\begin{document}

\title{Insights on Dark Matter from Hydrogen during Cosmic Dawn}

\author{Julian B.~Mu\~noz\footnote{Electronic address: \tt julianmunoz@fas.harvard.edu}
} 
\affiliation{Department of Physics, Harvard University, 17 Oxford St., Cambridge, MA 02138}
\author{Abraham Loeb}
\affiliation{Astronomy Department, Harvard University, 60 Garden St., Cambridge, MA 02138}

\date{\today}

\begin{abstract}
The origin and composition of the cosmological dark matter remain a mystery.
However, upcoming 21-cm measurements during 
cosmic dawn, the period of the first stellar formation,
can provide new clues on the nature of dark matter.
During this era, the baryon-dark matter fluid is the slowest it will ever be, making it ideal to search for dark matter elastically scattering with baryons through massless mediators, such as the photon.
Here we explore whether dark-matter particles with an electric ``minicharge" can significantly alter the baryonic temperature and, thus, affect 21-cm observations.
We find that the entirety of the dark matter cannot be minicharged at a significant level, lest it interferes with Galactic and extragalactic magnetic fields.
However, if minicharged particles comprise a subpercent fraction of the dark matter, and have charges $\epsilon \sim 10^{-6}$---in units of the electron charge---and masses $m_\chi \sim 1-60$ MeV, they can significantly cool down the baryonic fluid, and be discovered in 21-cm experiments.
We show how this scenario can explain the recent result by the EDGES collaboration, which requires a lower baryonic temperature than possible within the standard model, while remaining consistent with all current observations.
\end{abstract}

\maketitle

The dynamics of our Universe is strongly influenced by the pervasive, albeit elusive, dark matter (DM), known to outweigh regular baryonic matter roughly five to one~\cite{Ade:2015xua,Rubin}. 
Despite this, little is known about its nature 
and composition~\cite{Peebles:2000yy,Bertone:2004pz}.
All the evidence for DM relies on its gravitational pull on baryons, and thus does not require any nongravitational interactions between the two fluids.
Nonetheless, some level of nongravitational interactions can naturally explain their comparable cosmic abundances~\cite{Kaplan:1991ah,Allahverdi:2013mza,Allahverdi:2017edd}, as well as perhaps solve the well-known small-scale problems in the pure cold-DM models~\cite{Weinberg:2013aya}.
As of now, a vast array of direct- and indirect-detection experiments~\cite{Akerib:2017kat,Tan:2016zwf,Amole:2016pye,TheFermi-LAT:2017vmf}, cosmological observations~\cite{Cirelli:2009bb,Dvorkin:2013cea,Ali-Haimoud:2015pwa,Munoz:2017qpy,Gluscevic:2017ywp}, and accelerator-based probes~\cite{Fox:2011fx,Aaboud:2016tnv}, have not been able to find conclusive evidence for nongravitational interactions between DM and baryons, placing
stringent constraints on the origin of the DM, 
and on its interactions with standard-model particles.

A novel arena on the search for DM-baryon interactions can be found at cosmic dawn.
During this era, the first stars were formed~\cite{Barkana:2000fd,loeb2013first}, which coupled the spin temperature of neutral hydrogen to its much-lower kinetic temperature~\cite{Wout,Field}, causing cosmic-microwave-background (CMB) photons with a local wavelength of 21 cm to be resonantly absorbed by the intervening neutral hydrogen.
Eventually, however, the baryonic gas was heated by stellar remnants (such as X-ray binaries), and the hydrogen spin temperature increased above that of the CMB, causing 21-cm emission~\cite{Furlanetto:2006jb,Pritchard:2006sq}.
The absorption era provides one of the lowest-velocity environments in our Universe, where DM interactions with the visible sector mediated by a massless field are expected to be most prominent~\cite{Munoz:2015bca,Tashiro:2014tsa,Barkana}.
We will use this insight to explore the possibility that the DM interacts with the standard model (SM) through a small ``minicharge" under the usual electromagnetic force.

We adopt as a benchmark the measurement of the EDGES collaboration~\cite{Bowman}, which---if confirmed---indicates that baryons have a temperature of $T_b\approx 4$ K at $z\approx17$, roughly half of its expected value.
This poses a problem to the standard cosmological model, as
astrophysical processes would produce baryonic heating, rather than cooling~\cite{Barkana}.
However, elastic DM-baryon scattering can produce thermalization between the cold DM and the baryons, thus explaining this ``cold-baryon" problem\footnote{After this manuscript was completed a different explanation was put forward in Ref.~\cite{Feng:2018rje}, where it was suggested that an exotic radio excess at high redshift could potentially bias 21-cm results.
}.
In this work we will explore the region of the DM charge-mass plane that can be probed by 21-cm observations, and compare with current constraints.
We will argue that the minicharges required to cool down the baryons would be too large to allow the entirety of the DM to be charged, as it would not behave as a cold collisionless fluid.
Nonetheless, we will show that if a fraction $f_{\rm dm}\lesssim 10^{-2}$ of the DM is composed of particles with charge $\epsilon \sim 10^{-6} $, and mass $m_\chi \lesssim 60 \,{\rm MeV} \times (f_{\rm dm}/10^{-2})$, the baryonic temperature can be reduced by a factor of two, while being consistent with all other constraints.
We will conclude with some remarks about the 21-cm fluctuations induced by these interactions.

Endowing the dark matter with a small electric charge has unique phenomenological consequences, as charged particles respond to background magnetic fields, which are rather common in astrophysical environments.
In Ref.~\cite{Chuzhoy:2008zy} it was argued that supernova shocks would eject all minicharged particles from the Galactic disk, while the Galactic magnetic field, known to extend beyond Galactic heights of 3 kpc~\cite{Jansson:2012pc}, would prevent their reentry.
Given that the dark-matter density within 1.5 kpc of the disk is in agreement with predictions~\cite{Bovy:2012tw}, 
we can conclude that not all DM can be evacuated from the disk,
and thus,  minicharged particles with charges larger than $\epsilon/m_\chi \gtrsim 5\cdot 10^{-16}\, \rm MeV^{-1}$~\cite{McDermott:2010pa} (barring some fraction near the edge of that constraint, which could diffuse back to the disk~\cite{SanchezSalcedo:2010ev}), are precluded from being the entirety of the cosmological DM.
Here, and throughout, we define the minicharge $\epsilon\equiv e_\chi/e$ in units of the electron charge $e$, where $e_\chi$ is the DM charge, and we work in natural units, with $\hbar = c = 1$.

Minicharged particles can, however, avoid disk ejection if they cool efficiently. 
Moreover, we estimate the Galactic magnetic-field energy density to be at least three orders of magnitude smaller than the DM kinetic energy density in the Solar vicinity. Thus, DM could in principle be able to breach through magnetic-field lines and reenter the disk (albeit altering the magnetic-field structure of our Galaxy).
We note, however, that independent constraints can be achieved by requiring the DM to not be trapped in coherent regions of magnetic field in galaxy clusters, which have typical correlation lengths $r_{\rm corr}\sim10$ kpc and field strengths $B\sim 5\,\mu$G~\cite{Clarke:2000bz}. 
This means that charged particles with charges larger than $\epsilon/m_\chi \gtrsim 3 \cdot 10^{-17} \, \rm MeV^{-1}$ would not be distributed as cold dark matter, but instead clump wherever magnetic fields are coherent (or, if the DM breached through the field lines, these regions should lose coherence).
Additional constraints can be derived through plasma effects in cluster collisions, such as the bullet cluster~\cite{Heikinheimo:2015kra}, as well as by requiring the minicharged particles to not diffuse within clusters~\cite{Kadota:2016tqq}, although simulations might be required to isolate these effects  from nonlinear gravity~\cite{Sepp:2016tfs}.

The constraints discussed thus far would not apply if only a fraction of the DM is charged, as the majority of DM would behave as expected.
Given that the local DM measurements are accurate within tens of percent, we will
focus on the possibility that the minicharged particles compose a small fraction $f_{\rm dm} \leq 0.1$ of the dark matter, while the rest of it is neutral.
This can be naturally achieved if DM forms ``dark atoms"~\cite{Kaplan:2009de,Cline:2012is}, and there is a free charged-DM fraction after its recombination~\cite{CyrRacine:2012fz}.
Nonetheless, we will posit no assumptions about the origin of the minicharged particles, and parametrize them through their mass $m_\chi$ and charge $\epsilon$.
In that case, the \emph {momentum-transfer} cross section between a minicharged particle and a target $t$ (electron or proton) is \cite{McDermott:2010pa,Dvorkin:2013cea}
\be
\bar{\sigma}_t = \dfrac{2\pi \alpha^2 \epsilon^2 \xi}{\mu_{\chi,t}^2 v^4},
\label{eq:sigma}
\ee
where $\mu_{\chi,t}$ is the DM-target reduced mass, $\alpha$ is the fine-structure constant, and $v$ is the relative velocity between the two particles.
We have defined the Debye logarithm $\xi$, which regulates the forward divergence of the momentum-transfer integral~\cite{McDermott:2010pa,Dvorkin:2013cea}.
This factor is roughly constant during the era of interest, so we will set it to
\be
\xi = \log\left( \dfrac{9 T_b^3}{4\pi \epsilon^2 \alpha^3 x_e n_H}\right) \approx 68 - 2 \log \left(\dfrac{\epsilon}{10^{-6}}\right),
\ee
where we adopt a fiducial baryonic temperature $T_b = 10$ K, and we evaluate the free-electron fraction $x_e$, and the number density $n_H$ of hydrogen nuclei, at redshift $z=20$.

The velocity behavior of the cross section in Eq.~\eqref{eq:sigma} is indicative of where these interactions will be most prominent: wherever the DM-baryon (DM-b) fluid is slowest. 
The relative velocity between the DM and baryons is not only determined by their thermal motion. 
Baryons are impeded to collapse until recombination, whereas the DM is not, causing a velocity difference between them~\cite{Tseliakhovich:2010bj}.
Assuming that the DM is not strongly coupled enough to dissipate this velocity prior to recombination,
this velocity shows acoustic oscillations on Mpc scales~\cite{Tseliakhovich:2010bj}, and has a root-mean-square (rms) value of $v_{\rm rms} = 29$ km s$^{-1}\times [(1+z)/(1+z_{\rm kin})]$ after kinetic decoupling, starting at  $z_{\rm kin} \approx 1010$.
In Ref.~\cite{Munoz:2015bca} it was shown that DM-b interactions cause a drag, $D (v_{\chi,b}) \equiv  d v_{\chi,b}/dt$, on the DM-b relative velocity $v_{\chi,b}$, which here we recast as
\be
D(v_{\chi,b})  =  \sum_{t=e,p} \bar \sigma_t \dfrac{m_\chi n_\chi + \rho_b}{m_\chi+m_t} \dfrac{\rho_t}{\rho_b}  \dfrac{F(r_t)}{v_{\chi,b}^2},
\label{eq:dragV}
\ee
where $f_{\rm He}\equiv n_{\rm He}/n_H \approx 0.08$ is the Helium fraction,
$m_t$ is the target mass, $\rho_b$ is the energy density of baryons, and $\rho_t=x_e m_t n_H$ is the energy density of target particles. This drag depends on
the minicharged-DM number density, given by $n_\chi=f_{\rm dm} \rho_d/m_\chi$, where $\rho_d$ is the (total) DM energy density ($\rho_d = \Omega_c (1+z)^3 \rho_{\rm crit}$).
Here we have defined the function
\be
F(r_t) = \text{Erf}\left(\frac{r_t}{\sqrt{2}}\right)-\sqrt{\frac{2}{\pi }} r_t e^{-r_t^2/2},
\ee
where $r_t\equiv v_{\chi,b}/u_{{\rm th},t}$, and the (iso)thermal sound speed of the DM-$t$ fluid is given by
\be
u_{{\rm th},t}= \sqrt{\dfrac{T_b}{m_t} +\dfrac{T_\chi}{m_\chi} },
\label{eq:uth}
\ee
where $T_\chi$ is the minicharged-DM temperature.
By comparing this sound speed with the relative velocity, we can see that, immediately after recombination (and prior to the X-ray heating of the baryons), the baryonic sound speed falls below the DM-proton relative velocity, making the DM-proton (albeit not the DM-electron) fluid ``supersonic".

In addition to this drag,
interactions between DM and baryons will tend to bring both fluids into thermal equilibrium.
This gives rise to a baryonic heating~\cite{Munoz:2015bca}
\ba
\dot Q_b =& n_\chi \dfrac{x_e}{1+f_{\rm He}} \sum_{t=e,p} \dfrac{m_\chi m_t}{(m_\chi+m_t)^2} \dfrac{\bar{\sigma}_t}{ u_{{\rm th},t}}\times \nonumber\\
 &  \times\left[ \sqrt{\dfrac{2}{\pi}} \dfrac{e^{-r_t^2/2} }{u_{{\rm th},t}^2}(T_\chi-T_b) +  m_\chi \dfrac{F(r_t)}{r_t}  \right].
 \label{eq:Qdotb}
\end{align}
Here we have included DM interactions with both protons and electrons, as the latter can dominate if the minicharged-DM fluid is not cold.
The DM heating can be found by symmetry, through $n_\chi \to n_H\times(1+f_{\rm He})$, $m_\chi \leftrightarrow m_t$, and $T_\chi \leftrightarrow T_b$.
Notice that $\dot Q_b$ can in principle change signs depending on $r_t$, as for $r_t\to 0$ (corresponding to $v_{\chi,t}\ll u_{{\rm th},t}$) only the temperature-dependent term survives, corresponding to the usual thermalization; whereas for $r_t\gg1$ (which implies $v_{\chi,b}\gg u_{{\rm th},t}$), the heating term proportional to $F(r_t)$ can dominate, converting the mechanical energy of the relative velocity into positive heat for both fluids.

Equipped with Eqs.~\eqref{eq:dragV} and~\eqref{eq:Qdotb}, we can now obtain the thermodynamical evolution of these two systems by following Ref.~\cite{Munoz:2015bca} and solving
\begin{subequations}
	\ba
	\dot T_b &= -2 H T_b + 2\dot Q_b/3 + \Gamma_C(T_\gamma-T_b)
		\label{eq:ODE1} \\
	\dot T_\chi &= -2 H T_\chi + 2\dot Q_\chi/3 \\
	\dot x_e &= -C \left[ n_H \mathcal A_B x_e^2 - 4 (1-x_e) \mathcal B_B e^{3E_0/(4 T_\gamma)} \right]\\
	\dot v_{\chi,b} &= -H v_{\chi,b} - D(v_{\chi,b}),
	\label{eq:ODE4}
\end{align}
\end{subequations}
where $H$ is the Hubble parameter at time $t$,
$C$ is the Peebles factor~\cite{Peebles:1968ja,AliHaimoud:2010ab,Chluba:2010ca}, $E_0$ is the ground-level energy of the hydrogen atom, and $\mathcal A_B$ and $\mathcal B_B$ are the effective recombination/reionization coefficients, obtained from Ref.~\cite{AliHaimoud:2010dx}.
We have ignored photoheating and recombination cooling~\cite{Ali-Haimoud:2013hpa}, as well as possible baryonic heating due to DM annihilations, if these were present~\cite{Slatyer:2012yq}.
Here $T_\gamma$ is the temperature of the CMB photons, and the Compton thermalization rate is
\be
\Gamma_C = \dfrac{8 \sigma_T a_r T_\gamma^4 x_e}{3 (1+f_{\rm He}) m_e},
\ee
where $\sigma_T$ is the Thomson cross section, and $a_r$ is the Stefan-Boltzmann constant~\cite{Ma:1995ey,AliHaimoud:2010ab}.

The EDGES measurement~\cite{Bowman} requires the baryon temperature to be $T_b \approx 4$ K at a central redshift $z_{\rm central}=17$, a factor of two smaller than the usual result.
In order to halve the baryonic temperature with DM-b interactions, we require $m_\chi < \bar \mu_b \, f_{\rm dm} \Omega_c/\Omega_b$, where $\bar \mu_b\approx 1.2 \,m_p$ is the mean molecular weight of baryons, due to simple equipartition.
Thus, we will only show results for $m_\chi \leq 6.2 \, {\rm GeV} \times f_{\rm dm}$.
Moreover, for illustration purposes, we note that transferring half of the baryonic thermal energy to the DM at $z=z_{\rm central}$ would induce a DM sound speed of 
\be
u_\chi^2 (z_{\rm central})= \dfrac{T_\chi}{m_\chi} \approx \dfrac{T_b}{\bar{\mu}_b} \dfrac{ \Omega_b}{f_{\rm dm} \Omega_c} \approx  \dfrac{ \left( 0.1 \,\rm km\, s^{-1}\right)^2}{f_{\rm dm}},
\label{eq:uchi}
\ee
which redshifts as $(1+z)$.
Interestingly, for $f_{\rm dm} \lesssim 0.2$, this would give the minicharged component of the DM a sound speed larger than that of protons, which then determines the value of $u_{{\rm th},p}$ in Eq.~\eqref{eq:uth}. Moreover,  we estimate that for $f_{\rm dm} \lesssim 10^{-2}$ the DM-e interactions can dominate over the DM-p, due to this velocity.

\begin{figure}[hbtp!]
	\caption{Regions of the minicharged-particle parameter space that can be explored by 21-cm observations, and current constraints.
	Each solid line represents the minicharge required to	
	reduce the baryonic temperature by a factor of two, if a fraction $f_{\rm dm}$ of the DM is minicharged, where the color scales from black to red as $f_{\rm dm}$ decreases.
	Each line ends at $m_\chi=6 \, {\rm GeV} \times f_{\rm dm}$, as heavier particles would not be able to produce enough cooling.
	We note that, if $f_{\rm dm}=1$, all charges in the plot are ruled out, as explained in the main text.	
	In hatched we show the regions excluded by supernova cooling from 1987A (blue)~\cite{Chang:2018rso} (updated from Ref.~\cite{Essig:2013lka}, shown for reference in blue short-dashed line), 
	from a change in BBN (orange)~\cite{Jaeckel:2010ni}, and constraints from the SLAC millicharge experiment (brown)~\cite{Prinz:1998ua}.
	The purple hatched region represents the region above which DM would have efficiently annihilated into dark photons (if present), and caused a change in $N_{\rm eff} \gtrsim 1$~\cite{Vogel:2013raa}.
	Finally, in green long-dashed line we show the minicharge required to obtain the right DM abundance, for $f_{\rm dm}=1$, computed with Eq.~\eqref{eq:ThRel}.
	}
	\hspace{-10mm}
	\includegraphics[width=0.52\textwidth]{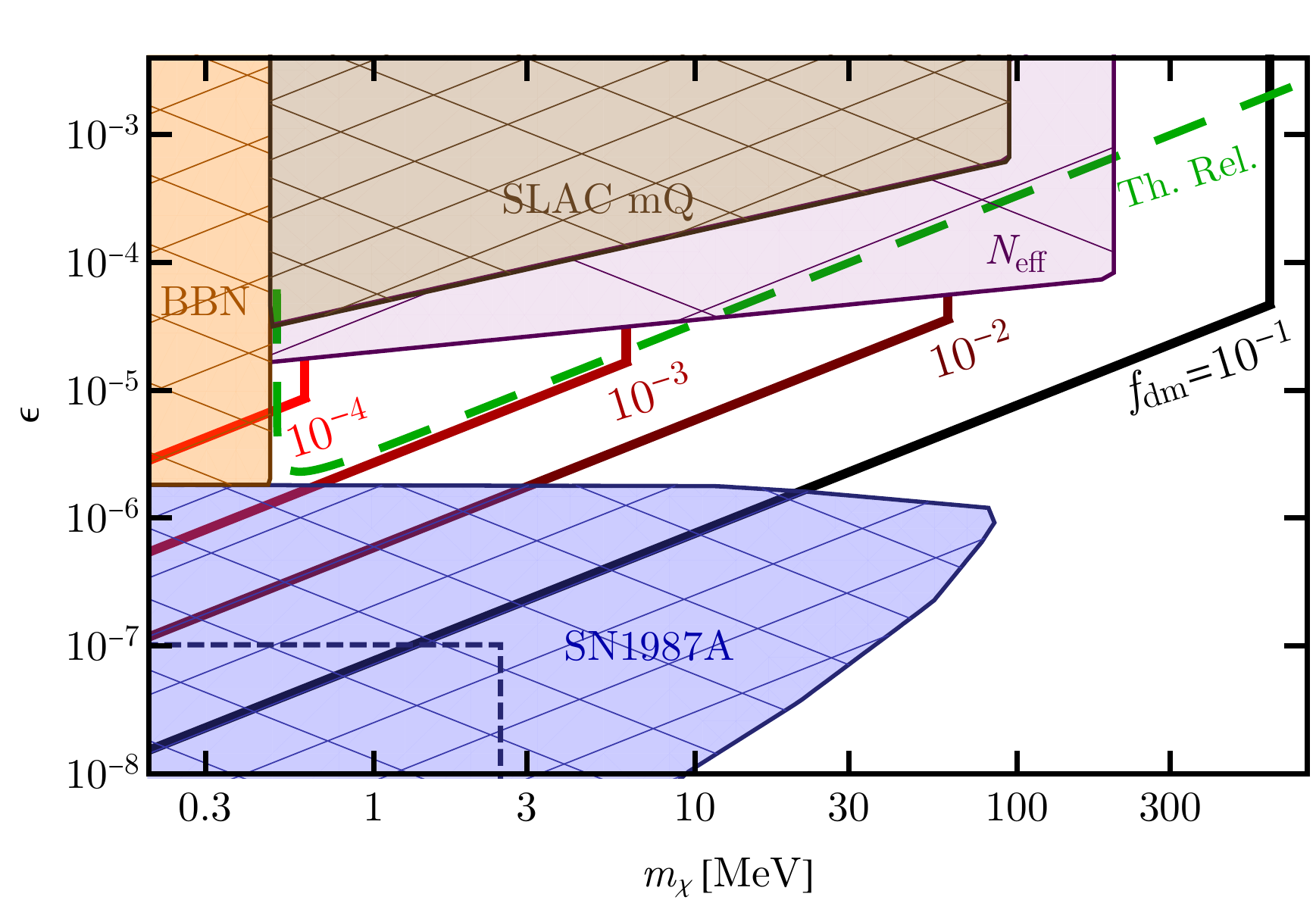}
	\label{fig:fdmrest}
\end{figure}

We solve the Eqs.~(\ref{eq:ODE1}-\ref{eq:ODE4}), starting at $z_{\rm kin}$, for different initial velocities $v_{\chi,b}^{(i)}$, and find $T_b(z,v_{\chi,b}^{(i)})$.
In order to remove dependencies on the coupling between the spin and kinetic temperatures, we define the average baryonic temperature as
\be
\VEV{T_b(z)} = \int dv_{\chi,b}^{(i)} \mathcal P(v_{\chi,b}^{(i)}) T_b(z,v_{\chi,b}^{(i)}),
\label{eq:Tbavg}
\ee
where the initial-velocity PDF is given by a Maxwell-Boltzmann distribution
\be
\mathcal P(v) =  \dfrac{4\pi v^2 e^{-3\,v^2/(2v_{\rm rms}^2)}}{(2 \pi v_{\rm rms}^2/3)^{3/2}},
\ee
with an rms velocity $v_{\rm rms} = 29$ km s$^{-1}$ at decoupling~\cite{Tseliakhovich:2010bj}.
We show, in Fig.~\ref{fig:fdmrest}, the line in the $\epsilon-m_\chi$ plane that would give a baryonic temperature of $\VEV{T_b}(z_{\rm central}) = 4$ K, for different values of $f_{\rm dm}$.
We find that, given $f_{\rm dm}$, the minicharge required always satisfies $\epsilon\propto m_\chi$ for $m_\chi < 6 \, {\rm GeV} \times f_{\rm dm}$. 
However, there is no simple analytic solution for the slope of this line as a function of $f_{\rm dm}$. This is because for small energy transfers the baryon heating is $\dot Q_b \propto f_{\rm dm} \epsilon^2/m_\chi^2$, whereas for large energy transfers (and assuming $f_{\rm dm}<0.2$), $\dot Q_b \propto f_{\rm dm}^{5/2} \epsilon^2/m_\chi^2$, given the $u_\chi$ scaling in Eq.~\eqref{eq:uchi}.
We have empirically found that 
\be
\epsilon(m_\chi, f_{\rm dm}) \approx 6 \cdot 10^{-7} \left( \dfrac{m_\chi}{\rm MeV}\right)\left( \dfrac{f_{\rm dm}}{10^{-2}}\right)^{-3/4},
\ee
is sufficient to reduce the baryonic temperature by a factor of two,
although we emphasize that the 21-cm results of Fig.~\ref{fig:fdmrest}  have been calculated numerically for each value of $f_{\rm dm}$.

We will now briefly review constraints on minicharged dark matter on the literature, to find how the 21-cm preferred regions in Fig.~\ref{fig:fdmrest} compare.

Minicharged particles lighter than $\sim100$ keV could be produced in stellar objects, such as white dwarfs and red giants, cooling these objects too rapidly~\cite{Davidson:2000hf}.
Similarly, minicharged-particle production during the supernova 1987A would have altered its neutrino luminosity, which places constraints in the range $2\times10^{-6}<\epsilon<10^{-9}$, where the upper limit is due to  self-absorption~\cite{Essig:2013lka,Chang:2018rso}. We label this limit as SN1987A in Fig.~\ref{fig:fdmrest}.

Current accelerators only limit minicharges $\epsilon\gtrsim 10^{-2}$~\cite{Davidson:1991si}, with the exception of the SLAC millicharge experiment, which constraints charges larger than $\epsilon\sim10^{-4}$ for $m_\chi<100\,\rm MeV$~\cite{Prinz:1998ua}. We show this constraint in Fig.~\ref{fig:fdmrest} labeled as SLAC mQ.
We note that a proposed extension to the LHC would allow it to probe the range $\epsilon \sim 10^{-3}$~\cite{Haas:2014dda}, for  the entire mass range we consider.

Measurements of the matter power spectrum, from the CMB and the Lyman-$\alpha$ forest, can constrain the DM minicharge to be $\epsilon < 2\cdot 10^{-9} \times (m_\chi/\rm MeV)$, for $f_{\rm dm}=1$ and DM masses between an MeV and a GeV~\cite{Dvorkin:2013cea} (see Ref.~\cite{Xu:2018efh} for an updated constraint).
However, if minicharged particles do not comprise all the DM these limits are less strict, and in particular, even particles with minicharges as large as $\epsilon \gtrsim 10^{-6} \times (m_\chi/\rm MeV)^{0.3}$, which would be in thermal contact with baryons at the CMB epoch, are allowed to compose up to a percent of the DM~\cite{Dolgov:2013una}.
This closes the apparent gap for $m_\chi \gtrsim 200$ MeV in Fig.~\ref{fig:fdmrest}, as the 21-cm data would require a fraction of the DM with charges above this threshold that is above a few percent.
Thus, we will focus on the $f_{\rm dm}<10^{-2}$ case for the rest of this Letter.

So far it has been sufficient for us to assume that minicharged particles compose a fraction $f_{\rm dm}$ of the dark matter, regardless of their origin.
However, the cosmology of minicharged particles can place constraints on their charge.
Particles with minicharges larger than $\epsilon\gtrsim 10^{-8} (m_\chi/\rm MeV)^{1/2}$, which encompasses the region of interest, would reach
equilibrium with the visible sector in the early Universe.
This severely limits minicharged particles lighter than electrons, as they would appear as additional light degrees of freedom ($N_{\rm eff}$) during big bang nucleosynthesis~\cite{Jaeckel:2010ni}. We label the constrained region as BBN in Fig.~\ref{fig:fdmrest}.
Moreover, if a dark photon ($\gamma'$) is present---as is expected to obtain minicharged particles~\cite{Holdom:1985ag}, although not necessary~\cite{Batell:2005wa}---this particle can also appear as light degrees of freedom during both BBN and the CMB~\cite{Vogel:2013raa,Foot:2014uba}.
We can estimate for what value of the minicharge $\epsilon$ dark photons would be produced,
by requiring that the timescale for two minicharged particles to annihilate into dark photons is longer than a Hubble time.
For minicharged particles in thermal equilibrium with the SM in the early universe, their
rate of annihilation into dark photons\footnote{Note that $\chi$ annihilations into $\gamma\gamma'$, or Compton-like processes ($\chi\gamma\to\chi\gamma'$) would be suppressed by a $\kappa^2$ factor, and at most alter the constraints by a $\mathcal O(1)$ factor.} is~\cite{Vogel:2013raa}
\be
\Gamma_{\chi \bar{\chi}\to \gamma'\gamma'} =  n_\chi \sigma v  \approx 10^{-3}  \, g'^4  T_\gamma ,
\ee
where $g'$ is the coupling constant between $\chi$ and $\gamma'$.
By requiring this rate to be smaller than $H \approx T_\gamma^2/M_{\rm pl}$, where $M_{\rm pl}$ is the reduced Planck mass, we can obtain a constraint on $g'$, so that DM does not annihilate to dark photons before $T_\gamma=m_\chi$.
Since the DM minicharge, $\epsilon$ is the product of the dark-photon mixing $\kappa$ times the dark coupling $g'$, and we require $\kappa<1$, this translates into the approximate constraint
\be
\epsilon \lesssim 2 \times 10^{-5} \left( \dfrac{m_\chi}{\rm MeV}\right)^{1/4},
\ee
for $m_\chi \leq \Lambda_{\rm QCD} \approx  200\,\rm MeV$, where $\Lambda_{\rm QCD}$ is the QCD scale,
which we label as $N_{\rm eff}$ in Fig.~\ref{fig:fdmrest}.
Here we have assumed that $\chi$ are spin-1/2 particles, and we note that this constraint can, of course, be tightened if $\kappa\ll1$, extending to $\chi$ masses as high as a GeV~\cite{Vogel:2013raa}.

In the standard freeze-out scenario, the DM production is halted when the baryonic temperature drops below its mass, and its annihilation rate determines the relic abundance left in the dark sector.
To compute the minicharge required to produce the right DM abundance, we use the approximate formula
\be
\Omega_c h^2 \approx 0.1 \times \left(\dfrac{x_f}{10}\right) \dfrac{10^{-26}\rm cm^3 s^{-1}}{(\sigma v)},
\label{eq:ThRel}
\ee
where $x_f=m_\chi/T_f$, $T_f$ is the freeze-out temperature, and
for minicharged dark matter the annihilation cross section to fermions is~\cite{McDermott:2010pa}
\be
(\sigma v) = \dfrac{\pi \alpha^2 \epsilon^2}{m_\chi^2} \sqrt{1-\dfrac{m_f^2}{m_\chi^2}} \left(1+\dfrac{m_f^2}{2 m_\chi^2}\right).
\ee
We will ignore any dark-sector interactions, and for simplicity, we will only consider annihilation into electron-positron pairs.
To obtain a simple estimate for this quantity
we further approximate $x_f$ to be a constant, as it only depends logarithmically on the DM mass and charge,
and find the region of the $\epsilon-m_\chi$ plane that produces the right DM relic abundance, which we show\footnote{We have assumed $f_{\rm dm}=1$ to obtain the thermal-relic line. To obtain a smaller fraction of DM with minicharges, the rest of it would have to form dark atoms, or otherwise have a larger minicharge.} in Fig.~\ref{fig:fdmrest}. A more detailed discussion can be found, for instance, in Refs.~\cite{McDermott:2010pa,Vogel:2013raa}.
From Fig.~\ref{fig:fdmrest} it is clear that---barring a small region for $m_\chi \sim$ few $\times$ MeV, and $\epsilon\gtrsim 10^{-6}$---most of the parameter space we are considering is below this thermal-relic line, implying that minicharged DM could not annihilate efficiently into SM particles, and would overclose the Universe.
Particles heavier than $\sim 200$ MeV could annihilate to light dark-sector fields, effectively producing a small $\Delta N_{\rm eff} \sim 0.1$~\cite{Brust:2013xpv,Green:2017ybv}. This small change of $N_{\rm eff}$ is below the sensitivity of current CMB probes, although within the reach of next-generation CMB experiments~\cite{Abazajian:2016yjj}.
However, for DM lighter than $\sim 200$ MeV, the SM bath has been heated by the QCD phase transition, and any populated light degrees of freedom in the dark sector will leave a trace on the CMB, as they cause $\Delta N_{\rm eff}\gtrsim1$ (depending on the nature of the light particles), strongly disfavored by Planck~\cite{Ade:2015xua}. 
In that case, other mechanisms may be invoked to set the right DM abundance~\cite{Griest:1990kh,DAgnolo:2017dbv,DAgnolo:2015ujb}, such as elastically decoupling relics~\cite{Kuflik:2015isi}, cannibal dark matter~\cite{Pappadopulo:2016pkp}, or having 3-to-2 annihilations dominate~\cite{Hochberg:2014dra}.
It is beyond the scope of this work to build a dark-sector model producing the right relic abundance, so we leave this question for future work.

Interestingly, minicharged particles can remain in the Galactic disk if they cool efficiently, for which minicharges $\epsilon>10^{-5} (m_\chi/\rm MeV)^{1/2}$ are required~\cite{Chuzhoy:2008zy}.
This might, nonetheless, not lead to direct-detection signals in, e.g., the Xenon~\cite{Essig:2012yx}, CRESST~\cite{Petricca:2017zdp}, or LUX~\cite{Dolan:2017xbu} experiments; 
as the gyroradius of these particles is $r_g \sim 10$ km $ (m_\chi/\rm MeV) \, (\epsilon/10^{-5})^{-1}$, on the terrestrial magnetic field of $B_\oplus \sim 0.1$ G.
They might, however, be detectable through atmospheric ionizations, acting as a \emph{nox borealis}, similar to the regular aurora borealis produced by solar-wind particles.
We point out, however, that Earth-based experiments could be sensitive to even a minuscule trace of minicharged particles, as these can interact rather strongly.
Given that neither disk ejection, due to the complex astrophysics of the interstellar medium; nor the terrestrial magnetic field would have perfect efficiency shielding the Earth from these particles, there might be hope for direct detection, especially through surface- and space-based experiments (as opposed to underground facilities). 
An example are the limits from the X-ray-calorimeter~\cite{Erickcek:2007jv}, which, however, do not constrain these particles if their masses are below $\sim 100\,\rm MeV$.
Additionally, we estimate that torsion-balance experiments~\cite{Wagner:2012ui}, which can constrain accelerations as small as $10^{-13}$  cm s$^{-2}$, could be sensitive to minicharges $\epsilon \sim 10^{-6} \, (m_\chi/\rm MeV)$, if the minicharged-particle density on the Earth's surface was one percent of the Solar-vicinity DM density.
This result is comparable to that required for baryonic cooling during cosmic dawn, although the specific number is controlled by the fraction of the minicharged-DM that would diffuse to Earth. More importantly, the cross section of these minicharged particles would be similar to the atmospheric column density, so any constraints would depend strongly on the DM momentum loss during atmospheric entry.

\begin{figure}[hbtp!]
	\hspace{-5mm}
	\includegraphics[width=0.46\textwidth]{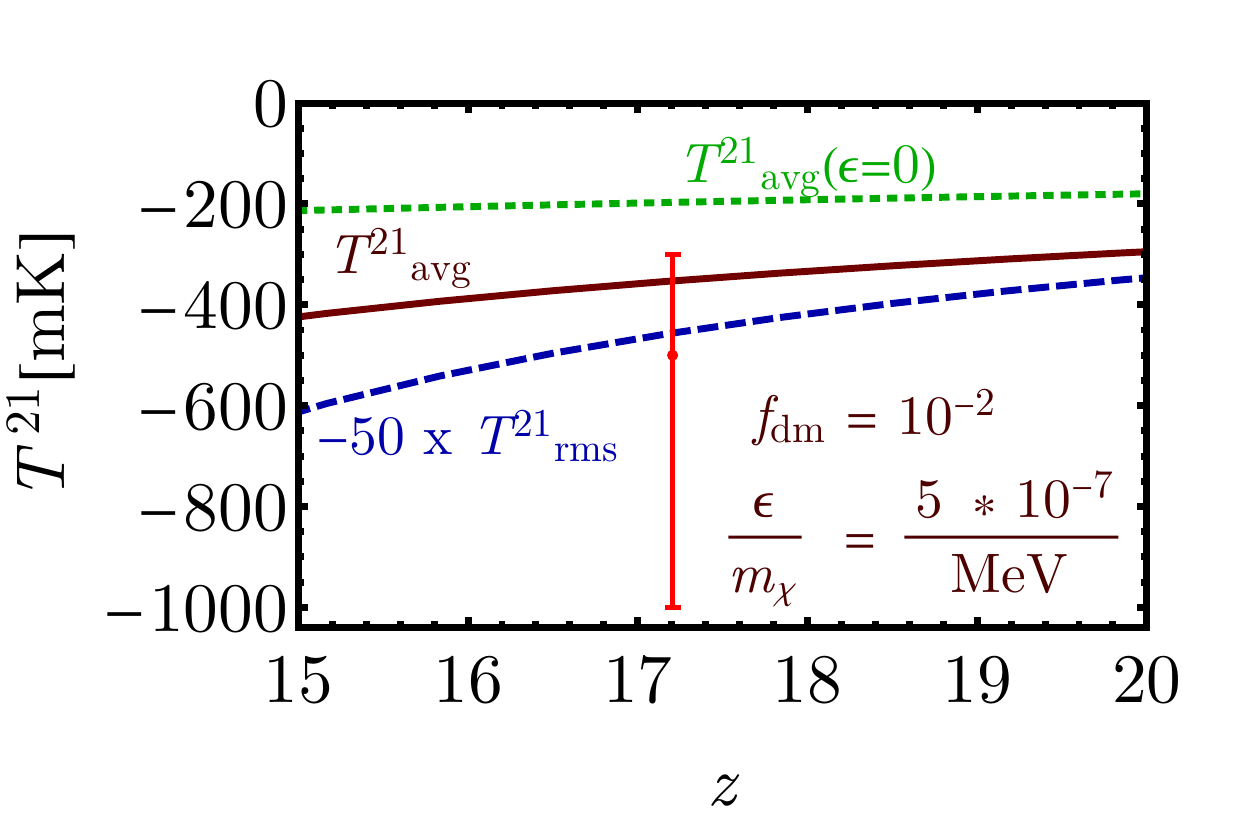}
	\caption{We show the 21-cm brightness temperature, obtained with Eq.~\eqref{eq:T21avg}, assuming full Lyman-$\alpha$ coupling and no X-ray heating, both in the case with (in red-solid line) and without (in green-dashed line) DM-b interactions, assuming that minicharged particles compose 1\% of the DM and have a charge $\epsilon = 5\, \cdot \, 10^{-7} \times (m_\chi/\rm MeV)$, with a mass $m_\chi \lesssim 60$ MeV.
	The red data point represents the data from EDGES~\cite{Bowman}, of $T^{(21)}=-500^{+200}_{-500}$ mK, at 3-$\sigma$.
	We also plot, in blue long-dashed line, the rms of the 21-cm temperature due to velocity fluctuations for this case, multiplied by a factor of $-50$.
	}
	\label{fig:T21}
\end{figure}

Let us now study how a change in baryonic temperature translates into an \emph{observable} 21-cm temperature.
The brightness temperature of the 21-cm line can be written as~\cite{Pritchard:2008da}
\be
T^{21} = 27\,{\rm mK} \, \left(\dfrac{T_s-T_\gamma}{T_s}\right)\left(\dfrac{x_{\rm HI} \Omega_b h^2}{0.023}\right)\left(\dfrac{0.15}{\Omega_m h^2} \dfrac{1+z}{10}\right)^{1/2}\!\!\!\!,
\ee
where $x_{\rm HI}$ is the neutral-hydrogen fraction (approximately unity during the region of interest), and
$T_s$ is the spin temperature of the hydrogen gas.
We will use the solutions for $T_b$ of Eqs.~(\ref{eq:ODE1}-\ref{eq:ODE4}), assuming full Ly-$\alpha$ coupling (so $T_s = T_b$)~\cite{Madau:1996cs}, to obtain the sky-averaged 21-cm temperature~\cite{Pritchard:2010pa,Cohen:2017xpx}
\be
T^{21}_{\rm avg} \equiv \VEV{T^{21}} = \int d v_{\chi,b} \mathcal P(v_{\chi,b}) T^{21}\left[T_b(v_{\chi,b})\right],
\label{eq:T21avg}
\ee
which we show in Fig.~\ref{fig:T21} for a specific choice of $\epsilon/m_\chi$, and $f_{\rm dm}$.
This Figure shows that the $T^{21}_{\rm avg}$ data from EDGES~\cite{Bowman} is in tension with the maximum absorption possible without DM-b interactions, whereas this tension is alleviated when introducing minicharged particles.

Moreover, DM-b interactions do not homogeneously cool down the baryons.
The DM-b relative velocity, with fluctuations over Mpc scales~\cite{Tseliakhovich:2010bj}, modulates the overall cooling/heating, thus sourcing additional 21-cm fluctuations~\cite{Munoz:2015bca}.
We can estimate the size of these fluctuations by finding the root mean square (rms) of the 21-cm brightness temperature, defined as
\be
T^{21}_{\rm rms} \equiv \sqrt{ \VEV{(T^{21})^2} - \VEV{T^{21}}^2}.
\label{eq:T21rms}
\ee
We show this function in Fig.~\ref{fig:T21}, where we can read that
the same interactions that cause a lower baryonic temperature also cause additional fluctuations, of amplitude $T^{21}_{\rm rms} \approx 10^{-2} \, T^{21}_{\rm avg}$ for $f_{\rm dm}=10^{-2}$.
These are comparable to the Mpc-scale adiabatic fluctuations at $z\sim 17$, of order $\delta_b (k=0.1\,  {\rm Mpc^{-1}},z=17) \sim 0.03$, thus perhaps making them detectable with the upcoming HERA experiment~\cite{DeBoer:2016tnn}, or the more-futuristic SKA~\cite{Mellema:2012ht}.
Notice, however, that even in the absence of interactions the DM-b relative velocity can affect the formation of the first structures in our Universe, and thus the 21-cm intensity~\cite{Fialkov:2011iw,Visbal:2012aw,Fialkov:2012su,Dalal:2010yt}.

Our results apply for minicharged particles interacting through any massive dark photon, as long it is lighter than the typical momentum transfer, which is $\sim \,\rm eV-keV$ for DM masses in the MeV$-$GeV range.
Additionally, we can easily translate our results for a DM minicharge to a new hadro/lepto-phillic DM interaction mediated by a light scalar $\phi$. For $f_{\rm dm}=1$ we found that a DM minicharge of $\epsilon \approx 10^{-8} \times (m_\chi/\rm MeV)$ is sufficient to decrease the baryonic temperature by a factor of two (although we remind the reader that this case is ruled out for minicharges).
Even ignoring DM self interactions~\cite{Harvey:2015hha}, and setting the $\phi$-DM coupling $g_\chi$ to unity,
the $\phi$-nucleon coupling required would be $g_N = 8\pi \alpha \epsilon (\bar x_e)^{-1/2} \approx 2\cdot10^{-11}\times (m_\chi/\rm MeV)$, where
$\bar x_e\approx 2\cdot 10^{-4}$ during the era of interest.
Similarly, the $\phi$-electron coupling required would be $g_e = 8\pi \alpha \epsilon (\bar x_e\,m_e/m_p)^{-1/2} \approx 10^{-9}\times (m_\chi/\rm MeV)$.
For DM in the MeV-GeV range (and thus mediators with $m_\phi\lesssim$ keV), this is constrained by stellar cooling, whereas for lighter dark matter the mediator would give rise to an anomalous fifth force \cite{Knapen:2017xzo,Stubbs}.

So far we have conservatively assumed that the small fraction of DM that is charged does not thermalize with the rest of the DM. If it did, one could simply rescale our results for $\epsilon$ from the $f_{\rm dm}=1$ case by $(f_{\rm dm})^{-1/2}$. Moreover, as a check, we have estimated the size of the interactions of minicharged particles with the neutral baryonic medium through Linhard's formula~\cite{Lindhard:1961zz,McDermott:2010pa}, and found that they are always subdominant, by at least four orders of magnitude.

In this Letter we have explored the possibility that part of the DM is charged under the usual electromagnetic force, albeit with a small minicharge $\epsilon$.
In that case, its momentum transfer with baryons is largest when their relative velocity is smallest, which occurs prior to reionization, during cosmic dawn.
We have shown that, while minicharged particles cannot comprise the majority of the DM, if a sub-percent fraction of the DM has a charge $\epsilon \sim 10^{-6}$, and mass $m_\chi \sim 1-60$ MeV, 
it can cause significant baryonic cooling, while being consistent with all current observations.
This cooling would lead to deeper 21-cm absorption, as recently reported by the EDGES collaboration~\cite{Bowman}, perhaps hinting at a dark-matter origin for the discrepancy.
We conclude that, through their 
access to the coldest epochs of our Universe, 21-cm experiments, such as EDGES~\cite{Bowman}, LEDA~\cite{LEDA}, and SARAS~\cite{Singh:2017syr},
provide us with a unique window into the dark sector, which may furnish the first nongravitational detection of the cosmological dark matter.

\acknowledgements

We wish to thank Prateek Agrawal, Yacine Ali-Haimoud, Cora Dvorkin, Marc Kamionkowski, Ely Kovetz, David Pinner, Christopher Stubbs, and Shawn Westerdale for enlightening discussions, and Hongwan Liu and Tracy Slatyer for finding a mistake in an earlier version of Eq.~\eqref{eq:dragV}.
This research is supported in part by the Black Hole Initiative, which is funded by a John Templeton Foundation grant.

\bibliography{Dmb_bib}{}

\begin{thebibliography}{10}

\bibitem{Ade:2015xua}
Planck, P.~A.~R. Ade {\em et~al.},
\newblock (2015), 1502.01589.

\bibitem{Rubin}
V.~C. {Rubin}, W.~K. {Ford}, Jr., and N.~{Thonnard},
\newblock ApJ {\bf 238}, 471 (1980).

\bibitem{Peebles:2000yy}
P.~J.~E. Peebles,
\newblock Astrophys. J. {\bf 534}, L127 (2000), astro-ph/0002495.

\bibitem{Bertone:2004pz}
G.~Bertone, D.~Hooper, and J.~Silk,
\newblock Phys. Rept. {\bf 405}, 279 (2005), hep-ph/0404175.

\bibitem{Kaplan:1991ah}
D.~B. Kaplan,
\newblock Phys. Rev. Lett. {\bf 68}, 741 (1992).

\bibitem{Allahverdi:2013mza}
R.~Allahverdi and B.~Dutta,
\newblock Phys. Rev. {\bf D88}, 023525 (2013), 1304.0711.

\bibitem{Allahverdi:2017edd}
R.~Allahverdi, P.~S.~B. Dev, and B.~Dutta,
\newblock Phys. Lett. {\bf B779}, 262 (2018), 1712.02713.

\bibitem{Weinberg:2013aya}
D.~H. Weinberg, J.~S. Bullock, F.~Governato, R.~Kuzio~de Naray, and A.~H.~G.
  Peter,
\newblock Proc. Nat. Acad. Sci. {\bf 112}, 12249 (2014), 1306.0913.

\bibitem{Akerib:2017kat}
LUX, D.~S. Akerib {\em et~al.},
\newblock Phys. Rev. Lett. {\bf 118}, 251302 (2017), 1705.03380.

\bibitem{Tan:2016zwf}
PandaX-II, A.~Tan {\em et~al.},
\newblock Phys. Rev. Lett. {\bf 117}, 121303 (2016), 1607.07400.

\bibitem{Amole:2016pye}
PICO, C.~Amole {\em et~al.},
\newblock Phys. Rev. {\bf D93}, 061101 (2016), 1601.03729.

\bibitem{TheFermi-LAT:2017vmf}
Fermi-LAT, M.~Ackermann {\em et~al.},
\newblock Astrophys. J. {\bf 840}, 43 (2017), 1704.03910.

\bibitem{Cirelli:2009bb}
M.~Cirelli, F.~Iocco, and P.~Panci,
\newblock JCAP {\bf 0910}, 009 (2009), 0907.0719.

\bibitem{Dvorkin:2013cea}
C.~Dvorkin, K.~Blum, and M.~Kamionkowski,
\newblock Phys. Rev. {\bf D89}, 023519 (2014), 1311.2937.

\bibitem{Ali-Haimoud:2015pwa}
Y.~Ali-Haïmoud, J.~Chluba, and M.~Kamionkowski,
\newblock Phys. Rev. Lett. {\bf 115}, 071304 (2015), 1506.04745.

\bibitem{Munoz:2017qpy}
J.~B. Muñoz and A.~Loeb,
\newblock JCAP {\bf 1711}, 043 (2017), 1708.08923.

\bibitem{Gluscevic:2017ywp}
V.~Gluscevic and K.~K. Boddy,
\newblock (2017), 1712.07133.

\bibitem{Fox:2011fx}
P.~J. Fox, R.~Harnik, J.~Kopp, and Y.~Tsai,
\newblock Phys. Rev. {\bf D84}, 014028 (2011), 1103.0240.

\bibitem{Aaboud:2016tnv}
ATLAS, M.~Aaboud {\em et~al.},
\newblock Phys. Rev. {\bf D94}, 032005 (2016), 1604.07773.

\bibitem{Barkana:2000fd}
R.~Barkana and A.~Loeb,
\newblock Phys. Rept. {\bf 349}, 125 (2001), astro-ph/0010468.

\bibitem{loeb2013first}
A.~Loeb and S.~R. Furlanetto,
\newblock {\em The first galaxies in the universe} (Princeton University Press,
  2013).

\bibitem{Wout}
S.~A. {Wouthuysen},
\newblock Astronomical Journal {\bf 57}, 31 (1952).

\bibitem{Field}
G.~B. {Field},
\newblock Astrophys. J. {\bf 129}, 536 (1959).

\bibitem{Furlanetto:2006jb}
S.~Furlanetto, S.~P. Oh, and F.~Briggs,
\newblock Phys. Rept. {\bf 433}, 181 (2006), astro-ph/0608032.

\bibitem{Pritchard:2006sq}
J.~R. Pritchard and S.~R. Furlanetto,
\newblock Mon. Not. Roy. Astron. Soc. {\bf 376}, 1680 (2007), astro-ph/0607234.

\bibitem{Munoz:2015bca}
J.~B. Muñoz, E.~D. Kovetz, and Y.~Ali-Haïmoud,
\newblock Phys. Rev. {\bf D92}, 083528 (2015), 1509.00029.

\bibitem{Tashiro:2014tsa}
H.~Tashiro, K.~Kadota, and J.~Silk,
\newblock Phys. Rev. {\bf D90}, 083522 (2014), 1408.2571.

\bibitem{Barkana}
R.~Barkana,
\newblock Nature {\bf To appear}.

\bibitem{Bowman}
J.~Bowman, A.~Rogers, R.~Monsalve, T.~Mozdzen, and N.~Mahesh,
\newblock Nature {\bf To appear}.

\bibitem{Feng:2018rje}
C.~Feng and G.~Holder,
\newblock (2018), 1802.07432.

\bibitem{Chuzhoy:2008zy}
L.~Chuzhoy and E.~W. Kolb,
\newblock JCAP {\bf 0907}, 014 (2009), 0809.0436.

\bibitem{Jansson:2012pc}
R.~Jansson and G.~R. Farrar,
\newblock Astrophys. J. {\bf 757}, 14 (2012), 1204.3662.

\bibitem{Bovy:2012tw}
J.~Bovy and S.~Tremaine,
\newblock Astrophys. J. {\bf 756}, 89 (2012), 1205.4033.

\bibitem{McDermott:2010pa}
S.~D. McDermott, H.-B. Yu, and K.~M. Zurek,
\newblock Phys. Rev. {\bf D83}, 063509 (2011), 1011.2907.

\bibitem{SanchezSalcedo:2010ev}
F.~J. Sanchez-Salcedo, E.~Martinez-Gomez, and J.~Magana,
\newblock JCAP {\bf 1002}, 031 (2010), 1002.3145.

\bibitem{Clarke:2000bz}
T.~E. Clarke, P.~P. Kronberg, and H.~Boehringer,
\newblock Astrophys. J. {\bf 547}, L111 (2001), astro-ph/0011281.

\bibitem{Heikinheimo:2015kra}
M.~Heikinheimo, M.~Raidal, C.~Spethmann, and H.~Veermäe,
\newblock Phys. Lett. {\bf B749}, 236 (2015), 1504.04371.

\bibitem{Kadota:2016tqq}
K.~Kadota, T.~Sekiguchi, and H.~Tashiro,
\newblock (2016), 1602.04009.

\bibitem{Sepp:2016tfs}
C.~Spethmann {\em et~al.},
\newblock Astron. Astrophys. {\bf 608}, A125 (2017), 1603.07324.

\bibitem{Kaplan:2009de}
D.~E. Kaplan, G.~Z. Krnjaic, K.~R. Rehermann, and C.~M. Wells,
\newblock JCAP {\bf 1005}, 021 (2010), 0909.0753.

\bibitem{Cline:2012is}
J.~M. Cline, Z.~Liu, and W.~Xue,
\newblock Phys. Rev. {\bf D85}, 101302 (2012), 1201.4858.

\bibitem{CyrRacine:2012fz}
F.-Y. Cyr-Racine and K.~Sigurdson,
\newblock Phys. Rev. {\bf D87}, 103515 (2013), 1209.5752.

\bibitem{Tseliakhovich:2010bj}
D.~Tseliakhovich and C.~Hirata,
\newblock Phys. Rev. {\bf D82}, 083520 (2010), 1005.2416.

\bibitem{Peebles:1968ja}
P.~J.~E. Peebles,
\newblock Astrophys. J. {\bf 153}, 1 (1968).

\bibitem{AliHaimoud:2010ab}
Y.~Ali-Haimoud and C.~M. Hirata,
\newblock Phys. Rev. {\bf D82}, 063521 (2010), 1006.1355.

\bibitem{Chluba:2010ca}
J.~Chluba and R.~M. Thomas,
\newblock Mon. Not. Roy. Astron. Soc. {\bf 412}, 748 (2011), 1010.3631.

\bibitem{AliHaimoud:2010dx}
Y.~Ali-Haimoud and C.~M. Hirata,
\newblock Phys. Rev. {\bf D83}, 043513 (2011), 1011.3758.

\bibitem{Ali-Haimoud:2013hpa}
Y.~Ali-Haïmoud, P.~D. Meerburg, and S.~Yuan,
\newblock Phys. Rev. {\bf D89}, 083506 (2014), 1312.4948.

\bibitem{Slatyer:2012yq}
T.~R. Slatyer,
\newblock Phys. Rev. {\bf D87}, 123513 (2013), 1211.0283.

\bibitem{Ma:1995ey}
C.-P. Ma and E.~Bertschinger,
\newblock Astrophys. J. {\bf 455}, 7 (1995), astro-ph/9506072.

\bibitem{Chang:2018rso}
J.~H. Chang, R.~Essig, and S.~D. McDermott,
\newblock (2018), 1803.00993.

\bibitem{Jaeckel:2010ni}
J.~Jaeckel and A.~Ringwald,
\newblock Ann. Rev. Nucl. Part. Sci. {\bf 60}, 405 (2010), 1002.0329.

\bibitem{Prinz:1998ua}
A.~A. Prinz {\em et~al.},
\newblock Phys. Rev. Lett. {\bf 81}, 1175 (1998), hep-ex/9804008.

\bibitem{Vogel:2013raa}
H.~Vogel and J.~Redondo,
\newblock JCAP {\bf 1402}, 029 (2014), 1311.2600.

\bibitem{Davidson:2000hf}
S.~Davidson, S.~Hannestad, and G.~Raffelt,
\newblock JHEP {\bf 05}, 003 (2000), hep-ph/0001179.

\bibitem{Essig:2013lka}
R.~Essig {\em et~al.},
\newblock {Working Group Report: New Light Weakly Coupled Particles},
\newblock 2013, 1311.0029.

\bibitem{Davidson:1991si}
S.~Davidson, B.~Campbell, and D.~C. Bailey,
\newblock Phys. Rev. {\bf D43}, 2314 (1991).

\bibitem{Haas:2014dda}
A.~Haas, C.~S. Hill, E.~Izaguirre, and I.~Yavin,
\newblock Phys. Lett. {\bf B746}, 117 (2015), 1410.6816.

\bibitem{Xu:2018efh}
W.~L. Xu, C.~Dvorkin, and A.~Chael,
\newblock (2018), 1802.06788.

\bibitem{Dolgov:2013una}
A.~D. Dolgov, S.~L. Dubovsky, G.~I. Rubtsov, and I.~I. Tkachev,
\newblock Phys. Rev. {\bf D88}, 117701 (2013), 1310.2376.

\bibitem{Holdom:1985ag}
B.~Holdom,
\newblock Phys. Lett. {\bf 166B}, 196 (1986).

\bibitem{Batell:2005wa}
B.~Batell and T.~Gherghetta,
\newblock Phys. Rev. {\bf D73}, 045016 (2006), hep-ph/0512356.

\bibitem{Foot:2014uba}
R.~Foot and S.~Vagnozzi,
\newblock Phys. Rev. {\bf D91}, 023512 (2015), 1409.7174.

\bibitem{Brust:2013xpv}
C.~Brust, D.~E. Kaplan, and M.~T. Walters,
\newblock JHEP {\bf 12}, 058 (2013), 1303.5379.

\bibitem{Green:2017ybv}
D.~Green and S.~Rajendran,
\newblock (2017), 1701.08750.

\bibitem{Abazajian:2016yjj}
CMB-S4, K.~N. Abazajian {\em et~al.},
\newblock (2016), 1610.02743.

\bibitem{Griest:1990kh}
K.~Griest and D.~Seckel,
\newblock Phys. Rev. {\bf D43}, 3191 (1991).

\bibitem{DAgnolo:2017dbv}
R.~T. D'Agnolo, D.~Pappadopulo, and J.~T. Ruderman,
\newblock Phys. Rev. Lett. {\bf 119}, 061102 (2017), 1705.08450.

\bibitem{DAgnolo:2015ujb}
R.~T. D'Agnolo and J.~T. Ruderman,
\newblock Phys. Rev. Lett. {\bf 115}, 061301 (2015), 1505.07107.

\bibitem{Kuflik:2015isi}
E.~Kuflik, M.~Perelstein, N.~R.-L. Lorier, and Y.-D. Tsai,
\newblock Phys. Rev. Lett. {\bf 116}, 221302 (2016), 1512.04545.

\bibitem{Pappadopulo:2016pkp}
D.~Pappadopulo, J.~T. Ruderman, and G.~Trevisan,
\newblock Phys. Rev. {\bf D94}, 035005 (2016), 1602.04219.

\bibitem{Hochberg:2014dra}
Y.~Hochberg, E.~Kuflik, T.~Volansky, and J.~G. Wacker,
\newblock Phys. Rev. Lett. {\bf 113}, 171301 (2014), 1402.5143.

\bibitem{Essig:2012yx}
R.~Essig, A.~Manalaysay, J.~Mardon, P.~Sorensen, and T.~Volansky,
\newblock Phys. Rev. Lett. {\bf 109}, 021301 (2012), 1206.2644.

\bibitem{Petricca:2017zdp}
CRESST, F.~Petricca {\em et~al.},
\newblock {First results on low-mass dark matter from the CRESST-III
  experiment},
\newblock 2017, 1711.07692.

\bibitem{Dolan:2017xbu}
M.~J. Dolan, F.~Kahlhoefer, and C.~McCabe,
\newblock (2017), 1711.09906.

\bibitem{Erickcek:2007jv}
A.~L. Erickcek, P.~J. Steinhardt, D.~McCammon, and P.~C. McGuire,
\newblock Phys. Rev. {\bf D76}, 042007 (2007), 0704.0794.

\bibitem{Wagner:2012ui}
T.~A. Wagner, S.~Schlamminger, J.~H. Gundlach, and E.~G. Adelberger,
\newblock Class. Quant. Grav. {\bf 29}, 184002 (2012), 1207.2442.

\bibitem{Pritchard:2008da}
J.~R. Pritchard and A.~Loeb,
\newblock Phys. Rev. {\bf D78}, 103511 (2008), 0802.2102.

\bibitem{Madau:1996cs}
P.~Madau, A.~Meiksin, and M.~J. Rees,
\newblock Astrophys. J. {\bf 475}, 429 (1997), astro-ph/9608010.

\bibitem{Pritchard:2010pa}
J.~R. Pritchard and A.~Loeb,
\newblock Phys. Rev. {\bf D82}, 023006 (2010), 1005.4057.

\bibitem{Cohen:2017xpx}
A.~Cohen, A.~Fialkov, and R.~Barkana,
\newblock (2017), 1709.02122.

\bibitem{DeBoer:2016tnn}
D.~R. DeBoer {\em et~al.},
\newblock Publ. Astron. Soc. Pac. {\bf 129}, 045001 (2017), 1606.07473.

\bibitem{Mellema:2012ht}
G.~Mellema {\em et~al.},
\newblock Exper. Astron. {\bf 36}, 235 (2013), 1210.0197.

\bibitem{Fialkov:2011iw}
A.~Fialkov, R.~Barkana, D.~Tseliakhovich, and C.~M. Hirata,
\newblock Mon. Not. Roy. Astron. Soc. {\bf 424}, 1335 (2012), 1110.2111.

\bibitem{Visbal:2012aw}
E.~Visbal, R.~Barkana, A.~Fialkov, D.~Tseliakhovich, and C.~Hirata,
\newblock Nature {\bf 487}, 70 (2012), 1201.1005.

\bibitem{Fialkov:2012su}
A.~Fialkov, R.~Barkana, E.~Visbal, D.~Tseliakhovich, and C.~M. Hirata,
\newblock Mon. Not. Roy. Astron. Soc. {\bf 432}, 2909 (2013), 1212.0513.

\bibitem{Dalal:2010yt}
N.~Dalal, U.-L. Pen, and U.~Seljak,
\newblock JCAP {\bf 1011}, 007 (2010), 1009.4704.

\bibitem{Harvey:2015hha}
D.~Harvey, R.~Massey, T.~Kitching, A.~Taylor, and E.~Tittley,
\newblock Science {\bf 347}, 1462 (2015), 1503.07675.

\bibitem{Knapen:2017xzo}
S.~Knapen, T.~Lin, and K.~M. Zurek,
\newblock (2017), 1709.07882.

\bibitem{Stubbs}
C.~W. Stubbs {\em et~al.},
\newblock Phys. Rev. Lett. {\bf 58}, 1070 (1987).

\bibitem{Lindhard:1961zz}
J.~Lindhard and M.~Scharff,
\newblock Phys. Rev. {\bf 124}, 128 (1961).

\bibitem{LEDA}
D.~C. {Price} {\em et~al.},
\newblock ArXiv e-prints  (2017), 1709.09313.

\bibitem{Singh:2017syr}
S.~Singh {\em et~al.},
\newblock (2017), 1710.01101.

\end{thebibliography}
\bibliographystyle{bibpreferences}

\end{document}